\documentclass[a4paper,fleqn,usenatbib]{mnras}
 
\usepackage{newtxtext,newtxmath}

\usepackage[T1]{fontenc}
\usepackage{ae,aecompl}
 
\usepackage{graphicx}
\usepackage{natbib}
\usepackage{amsmath}
\usepackage{lscape}
\usepackage{color}  
\usepackage{rotating}
\usepackage{booktabs}
\usepackage{caption}
\usepackage{ulem}
\usepackage{psfrag}
\usepackage{pstool}
\usepackage{bm}
\usepackage{mathrsfs}
\usepackage{mathtools}
 \usepackage{physics}
 
\providecommand{\av}[1]{\langle{#1}\rangle}
\providecommand{\aV}[1]{\left\langle{#1}\right\rangle}

\newsavebox{\astrutbox}
\sbox{\astrutbox}{\rule[-5pt]{0pt}{20pt}}

\def\curl{{\rm curl}\,}
\def\grad{{\rm grad}\,}
\def\div{{\rm div}\,}
\def\<{\langle}
\def\>{\rangle}

\def\bi{\bf}

\def\be{\begin{equation}}
\def\ee{\end{equation}}
\def\bea{\begin{eqnarray}}
\def\eea{\end{eqnarray}}
\def\nn{\nonumber}

\def\aap {{\it  Astron.\ Astrophys}. }

\def\apss{{\it  Astrophys.\ Space Sci}.}

\def\apj{{\it  Astrophys.\ J}. }
\def\apjs{{\it  Astrophys.\ J.\ Suppl}. }

\def\jgr{{\it  J.\ Geophys.\ Res}.\ }

\def\jpp{{\it  J.\ Plasma Phys}.}
\def\mnras{{\it  Mon.\ Not.\ R.\ Astron.\ Soc}. }

\newcommand{\br}{\null}

\title[A rotation-driven pulsar radio emission mechanism]{A rotation-driven pulsar radio emission mechanism}
 
\author[D. B. Melrose, M. Z. Rafat and A. Mastrano]%
{D. B. Melrose$^1$, M. Z. Rafat$^{1}$ and A. Mastrano$^{1}$\\
$^1$SIfA, School of Physics, University of Sydney, Sydney, NSW 2006, Australia}
 
\date{Accepted XXX. Received YYY; in original form ZZZ}
 
\pubyear{2017}
 
\begin{document}
\label{firstpage}
\pagerange{\pageref{firstpage}--\pageref{lastpage}}
\maketitle

\begin{abstract}
We propose and discuss an alternative pulsar radio emission mechanism that relies on rotation-driven plasma oscillations, rather than on a beam-driven instability, and suggest that it may be the generic radio emission mechanism for pulsars. We identify these oscillations as superluminal longitudinal waves in the pulsar plasma, and point out that these waves can escape directly in the O~mode. We argue that the frequency of the oscillations is $\omega_0\approx\omega_{\rm p}(2\langle\gamma\rangle)^{1/2}/\gamma_{\rm s}$, where $\gamma_{\rm s}$ is the Lorentz factor of bulk streaming motion and $\langle\gamma\rangle$ is the mean Lorentz factor in the rest frame of the plasma. The dependence of the plasma frequency $\omega_{\rm p}$ on radial distance implies a specific frequency-to-radius mapping, $\omega_0\propto r^{-3/2}$. Escape of the energy in these oscillations is possible if they are generated in overdense, field-aligned regions that we call fibers; the wave energy is initially refracted into underdense  regions between the fibers, which act as ducts. Some implications of the model for the interpretation of pulsar radio emission are discussed.
\end{abstract}
 
 
\section{Introduction}
 
In an accompanying paper  (Melrose, Rafat and Mastrano 2019, referred to here as Paper~1), we argue that none of the three most widely favored pulsar radio emission mechanisms is viable as a generic mechanisms for all pulsar radio emission: these are coherent curvature emission (CCE), relativistic plasma emission (RPE) and anomalous Doppler emission (ADE). It is highly desirable that an alternative radio emission mechanism be identified, one that avoids the severe constraints imposed by the assumption that the initial stage in the emission process is a resonant (e.g., beam-driven) instability. The emission mechanism we propose here is related to the electrodynamics of a rotating magnetic dipole. The possibility of such an alternative mechanisms has already been suggested in the literature \citep[e.g.,][]{B08,Lyubarsky09,T10b,TA13}:  large-amplitude oscillations (LAOs) are set up as the plasma attempts to screen the parallel component of the inductive electric field \citep{Letal05}. Our objective in this paper is to explore the suggestion that smaller-amplitude counterparts of such oscillations (SAOs) may lead directly to escaping radiation, and to propose this as the generic pulsar radio emission mechanism.
 
The assumption that pulsars are powered by rotational energy is widely accepted. In the forms of CCE, RPE and ADE discussed in Paper~1, the transfer of rotational energy is circuitous: first it is partly transferred to particles through acceleration by the rotation-induced parallel electric field, $E_\parallel$, with the accelerated particles emitting gamma-rays that decay into pairs, resulting in a pair cascade; beams of accelerated particles are assumed to form and to lead to growth of waves, which ultimately leads to the radio emission.  The waves that result  from beam-given wave growth are necessarily subluminal, defined here to mean phase speed $z=\omega/k_\parallel c<1$, where $ \omega $ is the frequency of emission and $ k_\parallel $ is the component of the wave vector $ \bm{k} $ parallel to the pulsar magnetic field. The difficulties identified in Paper~1 with beam-driven mechanisms are connected with the formation of beams and the very restricted range of subluminal phase speeds allowed for waves in a pulsar plasma. It is desirable to identify an emission mechanism that involves a more direct transfer of rotational energy and that avoids the limitations imposed by the requirement of beam formation and the restriction to subluminal phase speeds.

The idea that the radio emission can be driven directly by the rotational energy is not new, although the specific form discussed here is new. An earlier model is based on the suggestion that it is possible in principle for the centrifugal acceleration in a corotating magnetosphere to lead to an energy transfer to waves. Based on an idealized (bead-on-wire) model for the effect of the centrifugal force, \citet{MR94} argued that the motion of a particle on a corotating magnetic field line implies an oscillatory motion, which leads to a (parametric) plasma instability \citep{OMR02,MOM05,MRetal16}. The parametric instability is non-resonant, avoiding the restriction to subluminal phase speed. This parametric mechanism, which has also been suggested for active galactic nuclei \citep{GL97,RM00,RA08}. However, it has not received wide support as the basis for a pulsar radio emission mechanism \citep[e.g.,][]{2009arXiv0903.1113L}, and we do not discuss it further here. 

{\br
The emission mechanism proposed here involves a partial transfer of rotational energy to an oscillating electric field in the magnetosphere, with a fraction of the energy in the oscillating electric field escaping directly to produce the observed radio emission. The inductive electric field associated with an obliquely rotating magnetic dipole has a component, $E_\parallel$, parallel to the magnetic field. This $E_\parallel$ accelerates magnetospheric charges setting up a charge separation and associated potential electric field that tends to screen $E_\parallel$. The accelerated charges overshoot setting up large amplitude oscillations (LAOs) \citep[e.g.,][]{Letal05,BT07A,BT07,T10b}, as discussed in Section~\ref{sect:oscillations}. The importance of the intrinsically time-dependent nature of plasma discharges, {\br when considering the radio emission,} was pointed out by \cite{2020arXiv200102236P}, who proposed an emission mechanism similar to that proposed here.\footnote{The differences between these two mechanisms are discussed in Section~\ref{sect:discussion}.} Due to the intrinsic time dependence, it is not possible to screen $E_\parallel$ instantaneously everywhere. Any residual unscreened part may be regarded as small amplitude oscillations (SAOs). We argue that SAOs are widely distributed in the magnetosphere and that they produce the observed radio emission. }

The dispersive properties of the SAOs play an important role in the radio emission mechanism. In a nonrelativistic electron gas, such oscillations would be large-amplitude Langmuir waves, that is, large-amplitude longitudinal electron plasma oscillations at the plasma frequency, $\omega_p$. Assuming that the imposed electric field generates purely temporal oscillations that may be regarded a Langmuir waves with phase speed $z=\infty$. There are no Langmuir waves in a pulsar plasma, which we define as a one dimensional (1D) highly-relativistic, electron-positron plasma. The only longitudinal waves in a pulsar plasma are parallel-propagating L~mode waves. In the pulsar case, the waves generated in this way would correspond to  L-mode waves with $z=\infty$. A complication is that the plasma in a pulsar magnetosphere is streaming outward, and one needs to distinguish between the pulsar frame, ${\cal K}'$, in which $E_\parallel$ is varying purely in time, and the rest frame of the plasma, ${\cal K}$, in which the waves properties are determined. Waves generated at $z=\infty$ in one frame are propagating at a superluminal phase speeds, $z>1$, in any other frame.  The frequency of the L-mode waves, which plays an important role in the theory, depends on the phase speed. We assume that the SAOs are superluminal L-mode waves, and discuss how their frequency is determined separately.

An aspect of wave dispersion in a pulsar plasma that is directly relevant here is that L-mode waves can escape directly. The argument, cf. paper 1, is that the dispersion equation factors into one equation for the X-mode, which is strictly transverse and cannot be generated by longitudinal oscillations, and another equation for the two other modes. For parallel propagation, the other modes, the L-and A-modes say, may be regarded as counterparts of the Langmuir and Alfv\'en modes, respectively; the two dispersion curves cross at a cross-over point. {\br At the cross-over point, the L~mode dispersion curve is a decreasing function of $z$ and the Alfv\'en mode dispersion curve is $z=Z_{\rm A}$, with $z_{\rm A}=\beta_{\rm A}/(1+\beta_{\rm A}^2)^{1/2}\approx1-1/\beta_{\rm A}^2$, where $\beta_{\rm A}\gg1$ is the Alfv\'en velocity divided by $c$. Thus, the cross-over point is in the subluminal range. For slightly oblique propagation, the two modes reconnect, with the reconnected modes both avoiding the cross-over point. The oblique modes are the L-O-mode and a modified Alfv\'en mode. The L-O-mode is nearly longitudinal at frequencies well below the cross-over frequency, and nearly transverse at frequencies well above the cross-over frequency. As $\theta$ increases the L-O-mode dispersion curve moves to larger $z$, and at a tiny range of $\theta\approx0$ the O~mode dispersion curve crosses the light line and is entirely in the superluminal range for most $\theta$, as discussed in \S\ref{drO}. In principle, this allows the L-mode waves generated as SAOs to evolve into O-mode waves as they propagate along a prospective escape path. However, we argue that a specific type of plasma inhomogeneity is required to allow such escape to occur. Note that this conversion process produces purely (superluminal) O-mode emission. Although X-mode waves, which are subluminal, can escape, such waves would require a separate generation mechanism.}

In Section~\ref{sect:oscillations} we discuss the generation of the oscillations associated with incomplete screening of $E_\parallel$. In Section~\ref{sect:omega0} we appeal to the properties of wave dispersion in pulsar plasma to estimate the frequency, $\omega_0$, of the oscillations. In Section~\ref{sect:Jcor} we estimate the residual parallel electric field, resulting from incomplete screening, attributing it to a form of current starvation, and appeal to it in estimating the frequency spectrum of the resulting emission. In Section~\ref{sect:escape} we discuss propagation of the SAOs, emphasizing the role of refraction due to local density gradients in allowing radiation to escape. In Section~\ref{sect:discussion} we discuss the suggestion that all pulsar radio emission is due to this emission mechanism. We summarize our conclusions in Section~\ref{sect:conclusions}.

\section{Rotation-driven oscillations}
\label{sect:oscillations}
 
In this section we start by commenting on the fields around a rotating magnetic dipole, and then discuss the development of rotation-driven, nearly-temporal oscillations due to the partial screening of the parallel component, $E_{\rm ind\parallel}$, of the inductive electric field.

\subsection{Screening of $E_\parallel$}
\label{sect:screening}
 
The electromagnetic field around an obliquely-rotating magnetic dipole, $\bm{m}$, may be separated into three magnetic components, $ \bm{B} = \bm{B}_{\rm dip} + \bm{B}_{\rm ind} + \bm{B}_{\rm rad} $, and two electric components, $ \bm{E} = \bm{E}_{\rm ind} + \bm{E}_{\rm rad} $. The magnetic fields are its dipolar, $\abs{\bm{B}_{\rm dip}}\propto1/r^3$, inductive, $\abs{\bm{B}_{\rm ind}}\propto1/r^2$, and radiative, $\abs{\bm{B}_{\rm rad}}\propto1/r$, components, and the electric fields are its inductive, $\abs{\bm{E}_{\rm ind}}\propto1/r^2$, and radiative, $\abs{\bm{E}_{\rm rad}}\propto1/r$, components. The radiative components, for both $ \bm{B} $ and $ \bm{E} $, dominate at $r/r_{\rm L}\gg1$, where $ r_{\rm L} = cP/2\pi $ is the light cylinder radius. We are interested in regions $r\ll r_{\rm L}$, where only the leading components, ${\bi B}_{\rm dip}$ and ${\bi E}_{\rm ind}$ are important.. 
 
The parallel component $E_{\parallel\rm ind}=\hat{\bm{b}}\cdot\bm{E}_{\rm ind}$ along $\hat{\bm{b}}=\bm{B}/\abs{\bm{B}}\approx \bm{B}_{\rm dip}/\abs{\bm{B}_{\rm dip}}$, for $ r/r_{\rm L} \ll 1 $, is nonzero and changing as a function of time at the pulsar rotation frequency, $\omega_*=2\pi/P$, where $P$ is the period of the pulsar. Except when $E_{\rm ind\parallel}$ is perfectly screened, it accelerates electrons and positrons in opposite directions causing the parallel current, $J_\parallel = \hat{\bm{b}}\cdot\bm{J} $, to change as a function of time. Screening of an inductive field, $ \bm{E}_{\rm ind} $, by a potential electric field, $\bm{E}_{\rm pot}$, due to a charge density $\rho=\varepsilon_0\nabla\cdot\bm{E}_{\rm pot}$, is not possible in principle,\footnote{For example, the integral of the inductive electric field around any closed path is nonzero, and is unchanged by screening due to charges, which produce a potential field for which the integral around the closed path is zero.} but it is possible to screen the parallel component $E_{\rm ind\parallel}$. Perfect screening corresponds to the parallel components being equal and opposite $E_{\rm pot\parallel}+E_{\rm ind\parallel}=0$. The sum of the two electric fields is then equal to the corotation field, $\bm{E}_{\rm ind}+\bm{E}_{\rm pot}=\bm{E}_{\rm cor}=-(\bm{\omega}_*\times\bm{x})\times\bm{B}$, where $ \bm{x} $ is the position vector. Corotation requires a nonzero charge density $\rho_{\rm cor}=\varepsilon_0\nabla\cdot\bm{E}_{\rm cor}$. 
 
Any mismatch, $\rho-\rho_{\rm cor}\ne0$, between the actual, $ \rho $, and corotation, $ \rho_{\rm cor} $, charge densities, implies $\bm{E}-\bm{E}_{\rm cor}\ne0$, where $ \bm{E} $ is the actual electric field, and the parallel component of this mismatch field tends to drive (longitudinal) oscillations at the relevant natural frequency in the plasma. A temporally-dependent mismatch is driven by the time-varying inductive electric field. Such primarily temporal oscillations are the basis of the emission mechanism proposed here.
 
\subsection{Temporal oscillations}
\label{sect:temporal}
 
In the early literature, the formation of a stationary region with $E_\parallel\ne0$ was attributed to the fact that the corotation requirement $\rho=\rho_{\rm cor}$ cannot be satisfied everywhere along a field line due to charges originating from the stellar surface. The region with $E_\parallel\ne0$ was assumed to be confined to a stationary pair formation front (PFF), with the pairs created in the PFF screening $E_\parallel$ above the PFF. In the stationary case, electrodynamics reduces to electrostatics, and a model for a PFF is based on a 1D version of Gauss' equation. Similarly, for an oblique rotator  $\rho=\rho_{\rm cor}$ cannot be satisfied everywhere, and the mismatch $\rho-\rho_{\rm cor} \neq 0 $ becomes time dependent, with $\rho_{\rm cor}$ changing periodically at the rotation frequency $\omega_*=2\pi/P$. 
 
\subsubsection*{Large-amplitude oscillations (LAOs)}
 
The 1D forms of Amp\`ere's equation \citep{Letal05,BT07A,BT07,T10b} and Gauss' equation are, respectively,
\be
\frac{\partial E_\parallel}{\partial t}=-\frac{ J_\parallel-J_0}{\varepsilon_0},
\qquad 
\frac{\partial E_\parallel}{\partial s}=\frac{\rho-\rho_{\rm cor}}{\varepsilon_0},
\label{Levinson1}
\ee
where $s$ denotes distance along the magnetic field line. The term $J_0$ arises from the parallel component of $\curl\bm{B}$ and is assumed to include an average part determined by the global requirements of the electrodynamics \citep{S97}.  The term $J_0$ is also assumed to include fluctuating terms that arise from time-dependent boundary conditions, as suggested by the derivation of the first of equations (\ref{Levinson1}) given by \citet{T10b}. The model based on equations (\ref{Levinson1}), complemented by a kinetic equation that describes acceleration by $E_\parallel$, leads to oscillations. LAOs correspond to oscillations with a large enough amplitude for the potential drop due to the oscillating $E_\parallel$ to exceed the threshold required to trigger a pair cascade. As with models based on PFFs, effective pair creation is assumed to be confined to local regions (``inner gap'', ``outer gap'', ``slot gap'', etc.). The mismatch $\rho\ne\rho_{\rm cor}$ may be regarded as driving the LAOs in these local regions, although the LAOs result from the interplay of all three equations. 
 
Numerical models leading to LAOs \citep{Letal05,BT07A,BT07,T10b} are based on (\ref{Levinson1})  with additional assumptions relating the current to the particles, and including the effect of pair creation. Although there are differences in the details, the models show that the amplitude of $E_\parallel$ builds up initially, reaching a stage where $E_\parallel$ develops into a LAO with pair creation limiting the amplitude; the LAOs reach a quasi-steady state in which the threshold for pair creation is marginally satisfied, allowing the difference $\rho-\rho_{\rm cor}$ to remain small. This state is only ``quasi-steady'' in the sense that ongoing oscillations occur in a manner described as a limit cycle by \citet{TA13}. 
 
\subsubsection*{Small-amplitude oscillations (SAOs)}
 
Here we are concerned with oscillations that develop more widely through the magnetosphere once this quasi-steady state is reached. We refer to these as small-amplitude oscillations (SAOs). Unlike LAOs, which we assume to develop in charge-starved regions where a LAO triggers a pair cascade to provide the additional charges needed for screening, we assume that SAOs develop in regions where the pulsar plasma is present, and we suggest below that they are associated with current starvation rather than charge starvation. The concept of current starvation was discussed by \cite{Usov94} in connection with GRBs, and \cite{MM96} discussed the idea in connection with a pulsar wind; \cite{Melatos97} extended current starvation to magnetospheres including relativistic streaming and Compton drag.

In formulating a model for SAOs we assume that the oscillating $E_\parallel$, $J_\parallel$ and $\rho$ are described by $\delta E_\parallel$, $\delta J_\parallel$ and $\delta\rho$, and we look for three equations relating these wave quantities. Two of these equations are modified forms of equations (\ref{Levinson1}) and the third equation needs to describe the response of the plasma that determines the dispersive properties of the oscillations, which we identify as L-mode waves.
 
We assume that equations (\ref{Levinson1}) are replaced by
\be
\frac{\partial\delta E_\parallel}{\partial t}=-\frac{\delta J_\parallel-\delta J_0}{\varepsilon_0},
\qquad 
\frac{\partial\delta E_\parallel}{\partial s}=\frac{\delta\rho}{\varepsilon_0},
\label{Levinson2}
\ee
to describe the oscillating quantities, with $\delta J_0$ a source term associated with time-dependent boundary conditions. Equations (\ref{Levinson2}) need to be complemented by an additional (1D) equation relating $\delta J_\parallel$ and $\delta E_\parallel$. 
 
\subsection{Response of the plasma}
\label{sect:response}
 
The additional relation between $\delta J_\parallel$ and $\delta E_\parallel$ describes the response of the plasma, which involves the relativistic plasma dispersion function (RPDF) $z^2W(z)$ \citep[e.g.,][denoted RMM1 here]{2019JPlPh..85f9003R}, where $z$ is the phase speed of the wave. The parallel response in the rest frame of the plasma reduces to
\be
\frac{\partial\delta J_\parallel}{\partial t}=\varepsilon_0\omega_0^2\delta E_\parallel,
\label{Lw1}
\ee
with $\omega_0^2=\omega_{\rm p}^2z^2W(z)$. As discussed in RMM1, the RPDF has a sharp peak at $z=z_{\rm m}$, $1-z_{\rm m}\ll1$, that has a major effect on resonant beam-driven instabilities. This peak is not directly relevant here, where we are interested in superluminal phase speeds, $z>1$.

Combining equations (\ref{Levinson2}) and (\ref{Lw1}), the equation satisfied by the oscillations becomes
\be
\left[\frac{\partial^2}{\partial t^2}+\omega_0^2\right]\delta E_\parallel=
\frac{1}{\varepsilon_0}\frac{\partial\delta J_0}{\partial t}.
\label {Lw2}
\ee
where the right hand term is regarded as the source term for the oscillations. A system described by equation (\ref{Lw2}) tends to oscillate at frequency $\omega=\omega_0$. By Fourier transforming equation (\ref{Lw2}), it is evident that the source term can drive such oscillations provided that the Fourier transform of $\partial\delta J_0/\partial t$ is nonzero for $\omega=\omega_0$. We assume that the driving term includes fluctuations in the boundary conditions, associated with the field-aligned currents to and from the stellar surface required to provide cross-field current closure, as discussed below.
 
\section{Characteristic frequency}
\label{sect:omega0}
 
The frequency $\omega_0$ in equation (\ref{Lw1}) plays a central role in the proposed emission mechanism. It is identified as the characteristic frequency of the SAOs and is equated to the frequency of the observed radio emission, subject to an important proviso: these identifications apply in appropriate inertial frames. In this section we estimate the characteristic frequency $\omega_0$ assuming that the SAOs satisfy the dispersion relation for superluminal L~waves in a pulsar plasma.
 
\subsection{Two inertial frames}
 
We assume that the radio emission originates somewhere along open field lines where the pulsar plasma is streaming outward at speed $\beta_{\rm s}$ with Lorentz factor $\gamma_{\rm s}=(1-\beta_{\rm s}^2)^{-1/2}\gg1$. Two inertial frames are relevant: the rest (unprimed) frame of the plasma ${\cal K}$, and the pulsar (primed) frame ${\cal K}'$ in which the plasma is streaming.  The wave dispersion is most easily treated in ${\cal K}$, where the dispersion relation for the L~mode is $\omega=\omega_{\rm L}(z)$ with $\omega_{\rm L}^2(z)=\omega_{\rm p}^2z^2W(z)$, where $\omega_{\rm p}$ is the plasma frequency (without any Lorentz factors) and $z^2W(z)$ is the RPDF in ${\cal K}$.  In Appendix~\ref{sect:superluminal} analytic approximations to $z^2W(z)$ for superluminal phases speeds are given for a plasma with $\langle\gamma\rangle\gg1$ in ${\cal K}$.  Over the superluminal range the RPDF decreases from $z^2W(z)\approx2\langle\gamma\rangle$ at $z^2=1$ to $z^2W(z)=\langle\gamma^{-3}\rangle$ at $z^2=\infty$, with $\langle\gamma^{-3}\rangle\approx1/\langle\gamma\rangle$ \citep{MG99}. {\br (The plausible range of $\langle\gamma\rangle$  corresponds to a J\"uttner distribution with $\rho$ between 0.1 and 1 in ${\cal K}$. In Appendix~\ref{sect:superluminal} we complement our expressions for $\langle\gamma\rangle\gg1$ by giving numerical values for the specific case $\rho=1$, corresponding to $\langle\gamma\rangle\approx1.7$.)}
The frequency range for superluminal L~waves in ${\cal K}$ is $\omega_1\ge\omega\ge\omega_{\rm x}$, with $\omega_1^2=\omega_{\rm L}^2(1)\approx2\langle\gamma\rangle\omega_{\rm p}^2$ and $\omega_{\rm x}^2\approx\omega_{\rm p}^2/\langle\gamma\rangle$. We treat the wave dispersion in ${\cal K}'$ by Lorentz transforming the wave properties from ${\cal K}$.

A Lorentz transformation between the two frames implies the relations
\be
\omega'=\gamma_{\rm s}(\omega+k_\parallel c\beta_{\rm s}),
\quad
k'_\parallel c=\gamma_{\rm s}(k_\parallel c+\omega\beta_{\rm s}),
\quad
k'_\perp=k_\perp,
\label{LT1}
\ee
where $k_\parallel$ and $k_\perp$ are the parallel and perpendicular wavenumbers, respectively. The phase speeds, $z=\omega/k_\parallel c$, $z'=\omega'/k'_\parallel c$, are related by
\be
z'=\frac{z+\beta_{\rm s}}{1+\beta_{\rm s} z},
\quad
z=\frac{z'-\beta_{\rm s}}{1-\beta_{\rm s} z'},
\quad
\frac{z+\beta_{\rm s}}{z}=\frac{1}{\gamma_{\rm s}^2}\frac{z'}{z'-\beta_{\rm s}}.
\label{LT2}
\ee
The ratio of the frequencies in the two frames is
\be
\frac{\omega'}{\omega}=\gamma_{\rm s}\frac{z+\beta_{\rm s}}{z}
=\frac{1}{\gamma_{\rm s}}\frac{z'}{z'-\beta_{\rm s}}.
\label{ratio}
\ee

\begin{figure}
\begin{center}
\psfragfig[width=0.8\columnwidth]{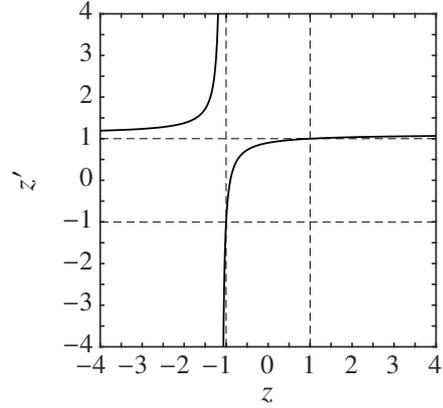}
 \caption{The relation (\ref{LT2}) between $z'$ and $z$ is plotted for $\beta_{\rm s}=0.9$, corresponding to $\gamma_{\rm s}\approx2.3$. The subluminal range corresponds to the curve inside the the box enclosed by the dashed lines at $z,z'=\pm1$. For a more realistic value of $\beta_{\rm s}$, corresponding to $\gamma_{\rm s}\gg1$, the curves approach straight lines at $z=-1$ and $z'=1$ except near where they would cross at $z=-z'=-1$.}
 \label{fig:zzp}  
 \end{center}
\end{figure}
 
We assume that the (superluminal) waves of interest are propagating outward, $z'>1$, in ${\cal K}'$. However, these may correspond to wave propagating either outward, $1<z<\infty$, or inward, $-\infty<z<-1/\beta_{\rm s}$, in ${\cal K}$. The relation between $z$ and $z'$, given by (\ref{LT2}), is plotted in Figure~\ref{fig:zzp}, which is for an artificially small value of  $\gamma_{\rm s}\approx2.3$. For larger values of $\gamma_{\rm s}$, the curves approach two straight lines at $z'=1$ and $z=-1$, deviating sharply away from each other near $(z,z') = (-1,1) $. In Figure~\ref{fig:zzp} the superluminal range is described by three separate sections, which correspond to $\{z,z'\}>1$, $\{z,z'\}<-1$ and $z<-1$ with $z'>1$, respectively. The third section of the superluminal waves in Figure~\ref{fig:zzp}, for $\gamma_{\rm s}\gg1$, separates into a nearly horizontal arm and a nearly vertical arm. 
 
\subsection{Nearly temporal oscillations in ${\cal K}$ and ${\cal K}'$}
 
Physical arguments are needed to determine the frequency range of the SAOs in ${\cal K}'$. Two limiting cases are strictly temporal oscillations in ${\cal K}$, corresponding to $k_\parallel=0$ or $z=\infty$, and to strictly temporal oscillations in ${\cal K}'$, corresponding to $k'_\parallel=0$ or $z'=\infty$.  We define ``nearly temporal oscillations'' in ${\cal K}$ and ${\cal K}'$, as the nearly horizontal arm and the nearly vertical arm, respectively, on the section with $z<-1$ with $z'>1$ in Figure~\ref{fig:zzp}. 
 
Strictly temporal oscillations in ${\cal K}$ have $\omega=\omega_{\rm x}$ in ${\cal K}$, and this transforms to $\omega' =\gamma_{\rm s}\omega_{\rm x}$ in ${\cal K}'$. Strictly temporal oscillations in ${\cal K}'$ have $z=-1/\beta_{\rm s}$ and $\omega=\omega_{\rm L}(1/\beta_{\rm s})\approx\omega_1$ in ${\cal K}$, and this transforms to $\omega'\approx\omega_1/\gamma_{\rm s}$ in ${\cal K}'$. The potential range of interest is between these two limiting cases. This range is $-\infty<z\le-1/\beta_{\rm s}$ and $\omega_{\rm x}<\omega\lesssim\omega_1$ in ${\cal K}$ and is $1/\beta_{\rm s}\le z'<\infty$ and $\gamma_{\rm s}\omega_{\rm x}<\omega'\lesssim\omega_1/\gamma_{\rm s}$ in ${\cal K}'$.
 
\subsection{Temporal oscillations in ${\cal K}'$}
 
Temporal oscillations in ${\cal K}$ might appear to be the most obvious generalization of the oscillations at $\omega_{\rm p}$ in a nonrelativistic plasma. (The frequency $\omega_{\rm x}$ is sometimes referred to as the relativistic plasma frequency.)
 However, there is a strong argument for the oscillations being nearly temporal oscillations in ${\cal K}'$: the driver is attributed to time-varying fields in the pulsar frame, that is in ${\cal K}'$. This driver is in ${\cal K}'$, leading one to expect that the temporal oscillations induced in the plasma have $k'_\parallel\approx0$, $z'\approx\infty$.
  
{\br A favorable condition for large wave growth is when the group speed of the waves is small, so that the waves experience many e-folding growths before propagating a distance over which the frequency changes as the plasma frequency charges.}

\subsection{Estimated value of $\omega_0$}

The assumption that SAOs correspond to nearly temporal oscillations with small group speed in ${\cal K}'$ implies that the frequency in ${\cal K}'$ is close to $\omega_1/\gamma_{\rm s}$. We assume that the frequency of SAOs, and hence of the escaping pulsar radio emission is 
\be
   \frac{\omega_0}{2\pi}{\br =\frac{(2\langle\gamma\rangle)^{1/2}\omega_p}{2\pi\gamma_{\rm s}} }
        \approx 0.4\zeta{\rm\,MHz},
\label{omega02a}
\ee
with
\be
\zeta=
        \left[\frac{\av{\gamma}}{10^2}\frac{\kappa}{10^5}
        \left(\frac{10^2}{\gamma_s}\right)^3
        \left(\frac{{\dot P}/P^3}{10^{-15}\,{\rm s}^{-3}}\right)^{1/2}\left(\frac{r/r_L}{0.1}\right)^{-3}\right]^{1/2}\left(\frac{1\,{\rm s}}{P}\right),
\label{omega02}
\ee
where we use the expression for $ \omega_{\rm p} $ from Paper~1.

\subsection{Parameters for observed range of frequencies}

\begin{figure}
\centering
\psfragfig[width=1.0\columnwidth]{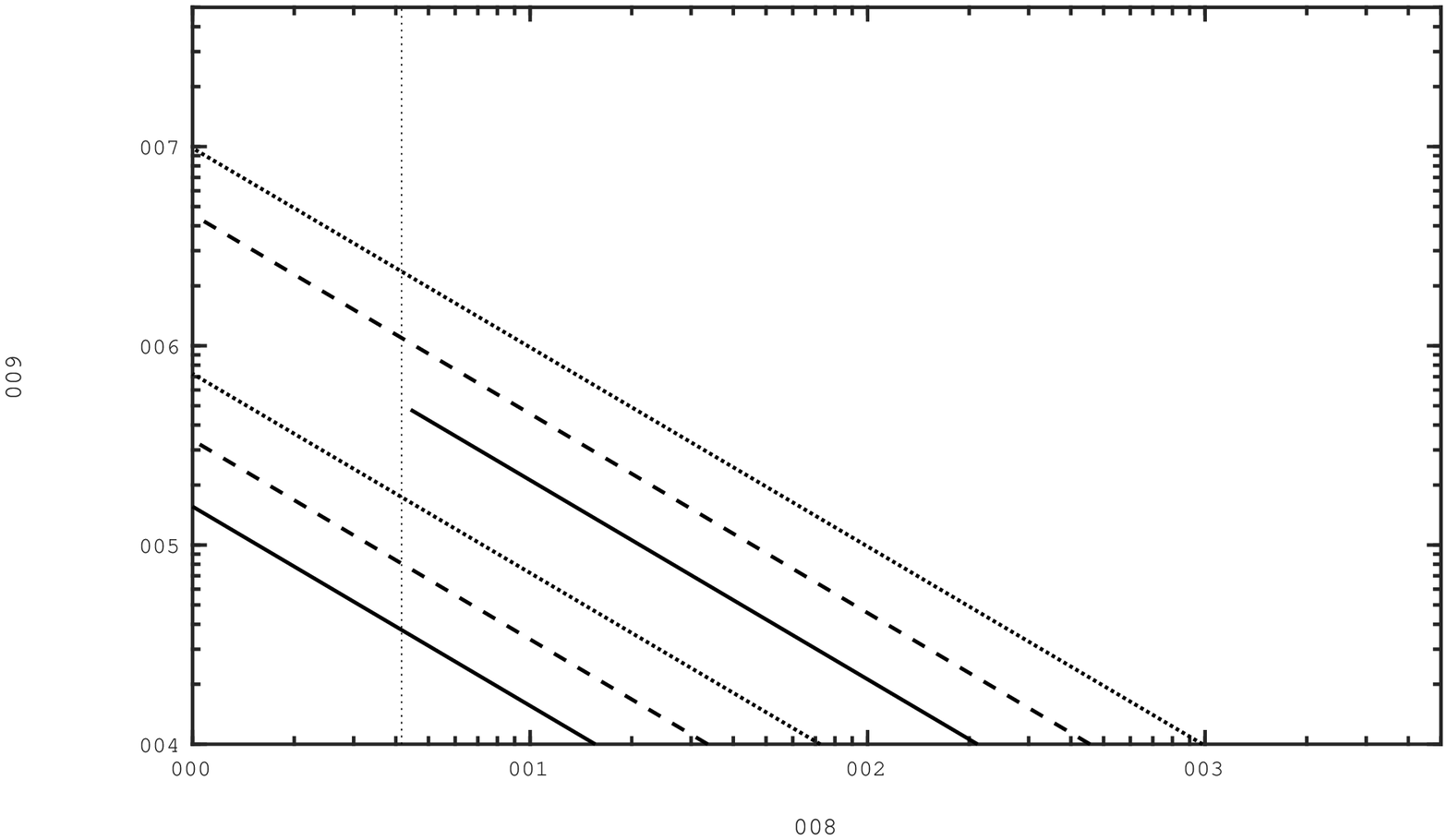}\\\vspace{-1mm}
\psfragfig[width=1.0\columnwidth]{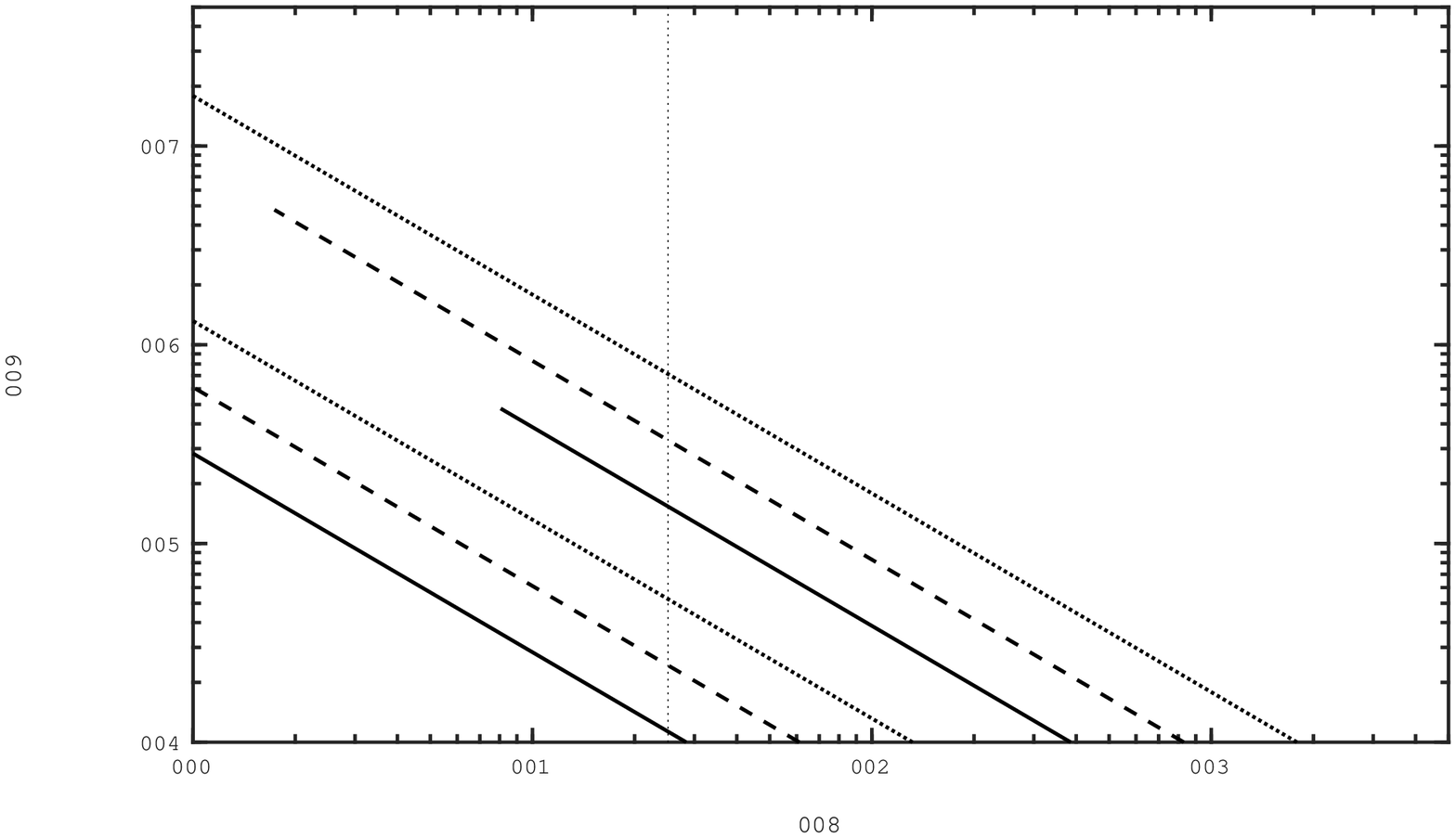}\\\vspace{-1mm}
\psfragfig[width=1.0\columnwidth]{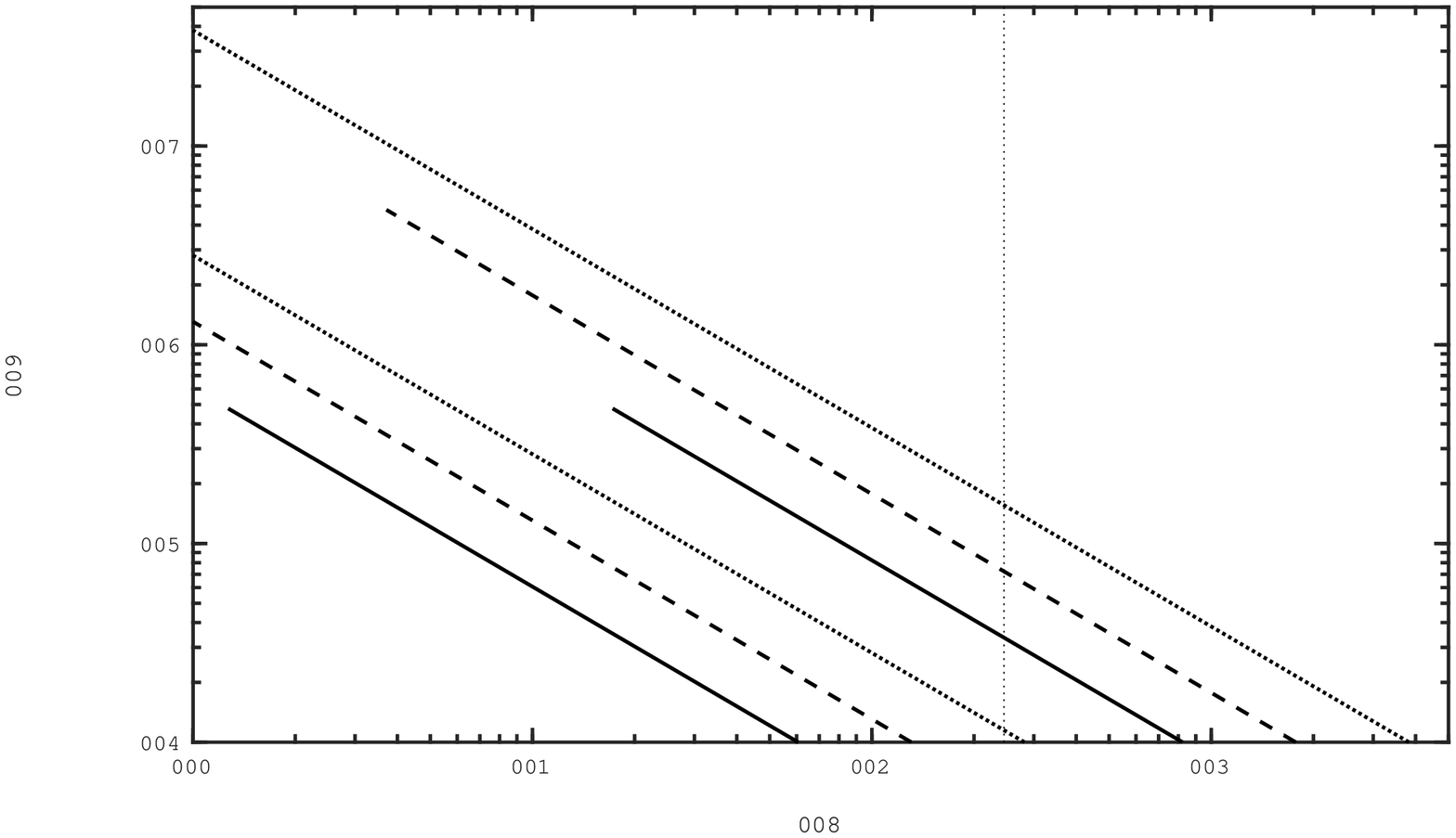}
\caption{Plots of $ \omega_0/2\pi $ corresponding to radio emission, $ 0.1\,{\rm GHz} \lesssim \omega_0/2\pi \lesssim 5\,{\rm GHz} $, for $ \av{\gamma} \approx 1.7 $ (TOP), 10 (MIDDLE) and 100 (BOTTOM); and $ P = 0.1\, {\rm s} $ (solid), 1 (dashed) and 10 (dotted). In each figure, each pair of lines correspond to 0.1\,GHz (upper) and 5\,GHz (lower); and the thin vertical dotted line corresponds to $ \gamma_{\rm s}^2 = 6\av{\gamma}^2 $. Each line runs from the stellar surface to $ r/r_L = 0.1 $. We use $ \kappa = 10^5 $ and $ \dot{P}/P^3 = 10^{-15}\,{\rm s}^{-3} $ for all plots.}
\label{fig:omega_0}  
\end{figure}

In Figure~\ref{fig:omega_0} we present plots of $ \omega_0/2\pi $, given by  (\ref{omega02a}) with (\ref{omega02}),  to illustrate the observed radio emission, which we assume to be $ 0.1\,{\rm GHz} \lesssim \omega_0/2\pi \lesssim 5\,{\rm GHz} $. We consider three values for $ \av{\gamma} \approx 1.7 $ (TOP), 10 (MIDDLE) and 100 (BOTTOM); we suggest that the plausible range is $1.7\lesssim\av{\gamma} \lesssim 10$, with the bottom panel corresponding to an extreme relativistic case. We present plots that cover the observed range of pulsar periods, from $ P = 0.1\, {\rm s} $ (solid), 1 (dashed) to 10 (dotted). In each figure, each pair of lines corresponds to 0.1\,GHz (upper) and 5\,GHz (lower). The thin vertical dotted line corresponds to $ \gamma_{\rm s}^2 = 6\av{\gamma}^2 $. Each line runs from the stellar surface to $ r/r_L = 0.1 $. In all cases we assume a multiplicity $ \kappa = 10^5 $ and $ \dot{P}/P^3 = 10^{-15}\,{\rm s}^{-3} $. 

 Although there is considerable uncertainty in the emission height, as discussed in Paper~1, most estimates are in the range from $ r/R_* \approx 4 $ to $ r/R_* \approx 200 $ (with a mean value of $ r/R_* = 20{\rm-}50 $) \citep{Mitra17}. In all cases, our model implies that the height increases with increasing pulsar period, consistent with observation \citep{Jetal08}. The height decreases with increasing $\gamma_{\rm s}$, and is nearly independent (increasing slowly) of $\langle\gamma\rangle$ for given $\gamma_{\rm s}$. Note however, that models for pair-cascades suggest a modest to large ratio $\gamma_{\rm s}/\langle\gamma\rangle$; for example, if one assumes $\gamma_{\rm s}=10\langle\gamma\rangle$, this fixes $\gamma_{\rm s}=17$, $10^2$ and $10^3$ in the three figures, respectively.

Based on these plots, we suggest that the most favorable cases are for small values of $\langle\gamma\rangle$ and modest values of $\gamma_{\rm s}/\langle\gamma\rangle$. For example, if the emission height is assumed very close to the stellar surface then $ \av{\gamma}=2$--3 and $ \gamma_{\rm s} \sim 10 $  would allow $ \omega_0/2\pi $ to account for pulsar radio emission. We do not discuss the details of possible choices here. A detailed discussion would need to include possible ranges of $\kappa$ and ${\dot P}/P^3$. {\br For example, our choice of $\kappa=10^5$ for a fiducial value is large, and $\kappa$ may be much smaller in some pulsars, reducing the value of $\omega_0$ correspondingly.} However, it is clear that there are plausible ranges of the various parameters for which the frequency in our model is compatible with the observed frequency range for all pulsars.

\section{Current starvation and $E_\parallel\ne0$}
\label{sect:Jcor}
 
In this section we outline an argument that implies $E_\parallel\ne0$ widely in the pulsar magnetosphere. We appeal to the value of $E_\parallel$ to estimate the maximum amplitude of the SAOs and the frequency spectrum of the radio emission.
 
\subsection{Corotation fields}
 
The corotation model for the electric field, charge and current densities for an obliquely corotating magnetosphere is summarized Section~\ref{sect:oscillations} and in Appendix~\ref{sect:A}. This corotation model leads to an inconsistency that we refer to as current starvation, implying that  $\bm{E}=\bm{E}_{\rm cor}$, $E_\parallel=0$ cannot be maintained instantaneously and continuously anywhere in the magnetosphere. The corotation charge density is determined by Gauss' equation and the corotation current density is determined by Amp\`ere's equation, cf.\ (\ref{Levinson1}). 
 
Amp\`ere's equation may be separated into three parts: a part, $\curl\bm{B}_0=\mu_0\bm{J}_0$ say, independent of the displacement current, a part $c^{2}\curl\bm{B}_{\rm ind}=\partial\bm{E}_{\rm ind}/\partial t$, that involves only the (inductive) fields generated by the rotating magnetic dipole, and a part $\bm{J}_{\rm cor}=-\varepsilon_0\partial\bm{E}_{\rm pot}/\partial t$, that involves only the fields generated by the charge density $\rho_{\rm cor}$. 
 
The corotation current density, $\bm{J}_{\rm cor}$, satisfies the continuity equation $\partial\rho_{\rm cor}/\partial t+\div\bm{J}_{\rm cor}=0$. We note that the current density, $\rho_{\rm cor}(\bm{\omega}_*\times\bm{x})$ associated with the rotating charge density, also satisfies this continuity equation. It follows that $\bm{J}_{\rm cor}$ may be written as
\be
\bm{J}_{\rm cor}=\rho_{\rm cor}(\bm{\omega}_*\times\bm{x})
+\bm{J}'_{\rm cor},
\label{Jcorp}
\ee 
with $\div\bm{J}'_{\rm cor}=0$. An explicit expression for $\bm{J}'_{\rm cor}$ is given in Appendix~\ref{sect:A}. An interpretation of (\ref{Jcorp}) is that $\bm{J}_{\rm cor}$ may be separated into two parts both of which correspond to closed current lines. The current lines corresponding to $\rho_{\rm cor}(\bm{\omega}_*\times\bm{x})$ are circles centered on the rotation axis; this part includes a steady current density $\propto m\cos\alpha(3\cos^2\theta-1)$ which produces a perturbation in the magnetic field equal to $\cos\alpha$ times the corresponding magnetic perturbation in the aligned case. The current lines of $\bm{J}'_{\rm cor}$ are also closed within the magnetosphere, with no sources or sinks. As a consequence, this current necessarily has components both along and across magnetic field lines.
 
\subsection{Current starvation}
 
The requirement that $\bm{J}'_{\rm cor}$ has a component across the magnetic field lines cannot be satisfied in general, invalidating the corotation model. Cross-field current flow in a magnetosphere may be attributed to drift motions, with the electric drift giving $\rho_{\rm cor}(\bm{\omega}_*\times\bm{x})$. However, the pressure-gradient and $\grad B$ drifts are zero in a pulsar plasma, due to the 1D motion of particles, and the curvature drift cannot produce an arbitrary cross-field current. We refer to the fact that the cross-field current density cannot be satisfied as current starvation.
 
Cross-field current closure is possible in another way: by the cross-field current in the magnetosphere being replaced by field-aligned current to and from the stellar surface, with cross-field closure there due to the finite conductivity. This possibility is familiar in the case of the Earth's magnetosphere, where it is referred to as a current wedge \citep{MRA73}, and also in the case of a laboratory plasma \citep{Simon55}. An additional feature in the pulsar case is that this form of closure introduces a time delay $\approx2r/c$ for $r\gg R_*$ \citep{MY16}, which corresponds to a (rotational) phase delay $\Delta\psi\approx2r/r_{\rm L}$ as this current attempts to maintain $\rho_{\rm cor}$ at its instantaneous (time-varying) value. (There is also a phase delay associated with the retarded time, $t-r/c$, which applies to all fields, and $\Delta\psi$ is an additional phase delay.) Qualitatively, this phase delay tends to smooth out the periodically varying part of $\rho_{\rm cor}$, cf. equation (\ref{dmf6a}), causing the local rotation velocity to differ from exact corotation, and leading to imperfect screening on $E_\parallel$.
 
\subsection{Force-free requirement}
 
The electromagnetic force density in the corotation model is $\rho_{\rm cor}\bm{E}_{\rm cor}+\bm{J}_{\rm cor}\times\bm{B}=\bm{J}'_{\rm cor}\times\bm{B}$. For the model to be valid, this force density must be balanced by some other force or by inertia. There is no other relevant force that can provide such balance. Let $\mu$ be the effective mass density; then the inertial force density in the model is of order $\mu r\omega_*^2$, where $r\omega_*^2$ is the centripetal acceleration. However, one finds that $\rho_{\rm cor}\bm{E}_{\rm cor}+\bm{J}_{\rm cor}\times\bm{B}$ is of order $\beta_{\rm A}^2$ times this inertial force density, with $\beta_{\rm A}^2\gg1$ in a pulsar magnetosphere (Paper~1). 
 
The force-free condition leads to a related inconsistency in the corotation model, and this may be regarded as an alternative argument for current starvation. The cross-field component of $\bm{J}'_{\rm cor}$ cannot be balanced, and hence cannot be maintained by the magnetospheric plasma. The current-wedge model for closure avoids this inconsistency, but introduces a phase delay.
 
\subsection{Phase delay due to current wedge}
 
The corotation electric field $\bm{E}_{\rm cor}$ may be written as the sum of the inductive electric field, $\bm{E}_{\rm ind}$, plus the potential field $\bm{E}_{\rm pot}$ due to $\rho_{\rm cor}$. In the absence of the phase delay due to the current wedge, one has $E_\parallel=E_{\rm ind\parallel}+E_{\rm pot\parallel}=0$. The phase delay does not affect the vacuum-field $\bm{E}_{\rm ind}$, but introduces a delay in the time-dependent part of $\rho_{\rm cor}$ and hence in $\bm{E}_{\rm pot}$.  The cancelation due to $E_{\rm pot\parallel}=-E_{\rm ind\parallel}$ is then not exact due to the phase delay of $E_{\rm pot\parallel}$ relative to $E_{\rm ind\parallel}$. The imbalance may be equated to the difference between $E_{\rm ind\parallel}$ at $\psi$ and at $\psi+\Delta\psi$. Let this difference be denoted by a tilde, so that one has
\be
{\tilde E}_\parallel=-\Delta\psi\frac{\partial}{\partial\psi}E_{\rm ind\parallel}.
\label{Epp1}
\ee
Using explicit expressions for $E_{\rm ind\parallel}$ and $J_{\rm cor\parallel}$ given by (\ref{Jcorp1}), equation~(\ref{Epp1}) gives
\be
{\tilde E}_\parallel=\Delta\psi\frac{J_{\rm cor\parallel}}{\varepsilon_0\omega_*},
\label{Epp2}
\ee
with $\Delta\psi\approx2r/r_{\rm L}$. An alternative expression is 
\be
{\tilde E}_\parallel=-\Delta\psi\frac{\mu_0m\omega_*}{4\pi r^2\Theta(\theta_{\rm b})}\sin\alpha\cos\alpha\,\sin\theta_{\rm b}\sin\phi_{\rm b},
\label{Epp3}
\ee
where the notation is defined in Appendix~\ref{sect:A}.

\subsection{Interpretation of ${\tilde E}_\parallel$}
 
The unbalanced parallel electric field ${\tilde E}_\parallel$ in the form (\ref{Epp3}) depends on the point in the magnetosphere, described by $r,\theta_{\rm b},\phi_{\rm b}$. The dependence on $\sin\phi_{\rm b}$ implies ${\tilde E}_\parallel=0$ for points in the plane containing the rotation and magnetic axes, with the maximum value at points orthogonal to this plane. The dependence on $\sin\theta_{\rm b}$ implies that the maximum value of ${\tilde E}_\parallel$ is at the magnetic equator, $\sin\theta_{\rm b}=1$. In the next section we consider the escape of radiation, and argue that this requires inhomogeneities that are plausible only on open field lines, that is, within the polar-cap region. Assuming that the last closed field line corresponds to $r=r_{\rm L}\sin^2\theta_{\rm b}$, the polar-cap region corresponds to $\sin\theta_{\rm b}\le(r/r_{\rm L})^{1/2}$. An estimate of the maximum value of ${\tilde E}_\parallel$, at the edge of the polar cap, is then
\be
{\tilde E}_\parallel=
-\left(\frac{r}{r_{\rm L}}\right)^{3/2}\frac{\mu_0m\omega_*}{4\pi r^2}\sin\alpha\cos\alpha\,\sin\phi_{\rm b},
\label{Epp4}
\ee
where we make the approximation $\Theta(\theta_{\rm b})\approx2$ for $\theta_{\rm b}\ll1$.
 
\subsection{Frequency spectrum of escaping radiation}
 
The energy density in the unbalanced parallel electric field is $\varepsilon_0{\tilde E}^2_\parallel/2$. We assume that this energy density provides an estimate of the energy density in SAOs driven in response to ${\tilde E}_\parallel$. A fraction of this energy can escape directly, as discussed in the next section. The maximum possible power, assuming all the SAOs escape, is then $c\varepsilon_0{\tilde E}^2_\parallel/2$ times the area of the source region. With $r_\perp=r\sin\theta_{\rm b}$ the cylindrical distance from the magnetic axis, we assume an annular area, $2\pi r_\perp\Delta r_\perp$, of width $\Delta r_\perp$ just inside the last closed field line. With $\sin\theta_{\rm b}\approx(r/r_{\rm L})^{1/2}$, this area corresponds to $3\pi r^2\Delta r/r_{\rm L}$. 
 
The frequency-to-radius relation $\omega\propto r^{-3/2}$ allows one to relate the power as a function of $r$ to the frequency spectrum. Assuming that the fraction that escapes is independent of $r$, this implies a maximum power $\propto r\Delta r\propto\omega^{-7/3}\Delta\omega$. This frequency spectrum is steeper than for pulsars on average. {\br If the fraction, $\mu(\omega)$ say, of the energy in SAOs that can escape is included, the power is multiplied by $\mu(\omega)$. This modified spectrum would reproduce the typical observed spectrum if $\mu(\omega)$ were an appropriately increasing function of $\omega$.} A detailed model is required to discuss this point further.

\section{Propagation of O~mode rays}
\label{sect:escape}
 
In this section we discuss the conditions under which SAOs initially in the L~mode can escape as O~mode waves. We start by summarizing analytic approximations for the O~mode dispersion relation in pulsar plasma, Hamilton's equations for a ray, and the constraint that the frequency $\omega'$ is constant in a frame in which the plasma is time-independent.
 
\subsection{Dispersion relation for the O~mode}
\label{drO}
 
The dispersion relation for the O~mode is ${\cal K}$ is (RMM1, with $b\approx1$)
\be
\omega^2=\omega_{\rm O}^2(z,\theta)=\frac{(z^2-z_{\rm A}^2)\,\omega_{\rm L}^2(z)}{z^2-z_{\rm A}^2-\tan^2\theta}.
\label{drO1}
\ee
For $\tan\theta=0$, this dispersion relation reduces to that for the L~mode, $\omega=\omega_{\rm L}(z)$ for $z^2>z_{\rm A}^2$, and to that for the parallel Alfv\'en mode, $z^2=z_{\rm A}^2$ for  $\omega>\omega_{\rm L}(z)$. For $\tan^2\theta\gtrsim1-z_{\rm A}^2\approx1/\beta_{\rm A}^2$, the O~mode is entirely superluminal.
 
It is helpful in the interpretation of the dispersion relation (\ref{drO1}) to compare it with the dispersion relation from transverse waves in an isotropic plasma. This dispersion relation may be written either as $n^2=1-\omega_{\rm p}^2/\omega^2$ or as $\omega^2=\omega_{\rm p}^2+k^2c^2$. When the latter form is rewritten in the notation used in (\ref{drO1}) it becomes $\omega^2=z^2\omega_{\rm p}^2/(z^2-1-\tan^2\theta)$. The denominator in this case is the same as the the denominator in (\ref{drO1}) in the limit $\beta_{\rm A}^2\to\infty$, $z_{\rm A}^2\to1$. Escape of transverse waves from an isotropic plasma, such as the solar corona when the magnetic field is neglected, requires that the frequency of the wave remain constant along a ray path over which $\omega_{\rm p}^2$ decreases to a negligible value. This requires that the denominator $z^2-1-\tan^2\theta$ decreases to a small value along this ray path in order for $\omega^2$ to remain constant. Similarly, for an O~mode wave to escape, the denominator in (\ref{drO1}) must become small along the ray path.
 
In discussing escape of waves, one needs to transform from ${\cal K}$ to the pulsar frame ${\cal K}'$. Using (\ref{ratio}), the dispersion relation (\ref{drO1}) transforms into 
\be
\omega'=\omega'_{\rm O}(z',\theta')=\gamma_{\rm s}\frac{z+\beta_{\rm s}}{z}\omega_{\rm O}(z,\theta),
\label{drO2}
\ee
with $z,\theta$ implicit functions of $z',\theta'$ on the right hand side, given by (\ref{LT2}) and
\be
\tan\theta'=\frac{\tan\theta}{\gamma_{\rm s}(1+\beta_{\rm s} z)},
\qquad
\tan\theta=\frac{\tan\theta'}{\gamma_{\rm s}(1-\beta_{\rm s} z')}.
\label{LT3}
\ee 
 
\subsection{Hamilton's equations for a ray}
 
In order to discuss how escape occurs we need equations that determine how $z'$ and $\theta'$ change along a ray path in the pulsar frame due to spatial gradients in $\omega_{\rm p}$. 
 
Hamilton's equations for a ray in ${\cal K}'$ are
\be
\frac{\dd s'}{\dd t}=\frac{\partial\omega'}{\partial k'_\parallel},
\quad
\frac{\dd x'}{\dd t}=\frac{\partial\omega'}{\partial k'_\perp},
\quad
\frac{\dd k'_\parallel}{\dd t}=-\frac{\partial\omega'}{\partial s'},
\quad
\frac{\dd k'_\perp}{\dd t}=-\frac{\partial\omega'}{\partial x'},
\label{HE1}
\ee
where $s'$ and $x'$ denote distance along and across (in the plane of the gradient) the magnetic field, respectively. We regard $\omega'$ as a function of $z',\theta',s',x'$, with $k'_\parallel=\omega'/cz'$, $k'_\perp=(\omega'/cz')\tan\theta'$. Then the first two of equations (\ref{HE1}) become
\be
\frac{\dd s'}{\dd t}=-\frac{cz'^2}{\omega'}\frac{\partial\omega'}{\partial z'},
\quad
\frac{\dd x'}{\dd t}=-\frac{cz'^2\cos\theta'}{\omega'\sin\theta'}\frac{\partial\omega'}{\partial z'}
+\frac{cz'\cos^2\theta'}{\omega'}\frac{\partial\omega'}{\partial\theta'},
\label{HE2}
\ee
with $\partial\omega'/\partial z'$ and $\partial\omega'/\partial\theta'$ found by differentiating equation (\ref{drO2}) with \eqref{drO1}. The remaining two of equations (\ref{HE1}) are replaced by
\be
\frac{\dd z'}{\dd t}=\frac{cz'^2}{\omega_{\rm p}}\frac{\partial\omega_{\rm p}}{\partial s'},
\qquad
\frac{\dd\theta'}{\dd t}=-\frac{cz'\cos\theta'}{\omega_{\rm p}}\frac{\partial\omega_{\rm p}}{\partial x'}
+\frac{cz'\sin\theta'}{\omega_{\rm p}}\frac{\partial\omega_{\rm p}}{\partial s'}.
\label{HE3}
\ee
 
Formally, $t$ in equations (\ref{HE1})--(\ref{HE3}) is an affine parameter that is sometimes interpreted as time. One may replace $t$ by the physically meaningful parameter $s'$ by dividing by $ds'/dt$ using the first of equations (\ref{HE1}), with the right-hand side interpreted as the group velocity. For propagation in the pulsar frame ${\cal K}'$ one has
\be
\frac{\dd z'}{\dd s'}\approx\frac{z'^2}{\beta'_{\rm g}\omega_{\rm p}}\frac{\partial\omega_{\rm p}}{\partial s'},
\qquad
\frac{\dd\theta'}{\dd s'}\approx-\frac{z'\cos\theta'}{\beta'_{\rm g}\omega_{\rm p}}\frac{\partial\omega_{\rm p}}{\partial x'}
+\frac{z'\sin\theta'}{\beta'_{\rm g}\omega_{\rm p}}\frac{\partial\omega_{\rm p}}{\partial s'}.
\label{HE5}
\ee
Due to the factors $1/\beta'_{\rm g}$, the effect of refraction is greatly enhanced in the neigborhood of the stationary-energy condition $\beta'_{\rm g}=0$, which corresponds to $\beta_{\rm g}=-\beta_{\rm s}$.

\subsection{Parallel-propagating waves cannot escape}

{\br Assuming the waves are generated as superluminal, parallel-propagating L-mode waves, they cannot escape directly if the only gradient in $\omega_{\rm p}$ is along the field lines ($\partial\omega_{\rm p}/\partial x'=0$). For an initially parallel-propagating wave to escape it must continue to propagate parallel to the field lines in the direction of decreasing $\omega_p$. The wave must also continue to satisfy the dispersion relation, which consists of two parts joined at the cross-over point: $\omega=\omega_{\rm L}(z)$ for $z>z_{\rm A}$ and $z=z_{\rm A}$ for $\omega=\omega_{\rm L}(z)$. A wave generated at $z'=\infty$ or $z^2=1/\beta^2_{\rm s}$ must propagate first to decreasing $z$ until it reaches $z=z_{\rm A}<1$. Although this seems possible, because $z^2W(z)$ is a monotonically increasing function of decreasing $z$ in this range and may balance the assumed decrease in $\omega_p$ to allow $\omega$ to remain constant, the detailed analysis below implies that it is possible only in two special cases. At the cross-over point the parallel O~mode becomes the parallel Alfv\'en mode, and to escape it must follow the dispersion curve $z=z_{\rm A}$ to arbitrarily large $\omega/\omega_{\rm p}$. }

The constant-frequency condition requires $\dd \omega'/\dd s'=0$ along the ray path. The spatial dependence of the dispersion relation, $\omega'^2=\omega_{\rm p}^2z'^2W'(z')=\omega_{\rm p}^2(z+\beta_{\rm s})^2W(z)$, includes the dependence of $\omega_{\rm p}$ on $s'$, described by the gradient $\partial\omega_{\rm p}/\partial s'$, and an implicit dependence through the spatial gradient in $z'$ along the ray path, determined by (\ref{HE5}). Together these imply, after some algebra,
\be
\frac{1}{\omega'}\frac{\dd \omega'}{\dd s'}=\frac{1}{\omega_{\rm p}}\frac{\partial\omega_{\rm p}}{\partial s'}
\frac{\beta'_{\rm g}}{\beta'_{\rm g}-z'}=\frac{1}{\omega_{\rm p}}\frac{\partial\omega_{\rm p}}{\partial s'}
\frac{\gamma_{\rm s}^2(\beta_{\rm g}+\beta_{\rm s})(1+z\beta_{\rm s})}{\beta_{\rm g}-z},
\label{dwpdsp}
\ee
where $\beta_{\rm g}$ and $\beta'_{\rm g}$ are the group speeds in ${\cal K}$ and ${\cal K}'$, respectively.
There are only two special cases where the frequency is constant. One case is $\beta'_{\rm g}=0$ or $\beta_{\rm g}=-\beta_{\rm s}$, which corresponds to the wave energy being stationary in ${\cal K}'$. In this case $\omega'$ does not change with $s'$ because there is no energy propagation (and hence no ray) in ${\cal K}'$. The other case is $z'=\infty$, $z=-1/\beta_{\rm s}$, that is, strictly temporal oscillations in ${\cal K}'$, corresponding to the transition between inward and outward propagating waves in ${\cal K}$. Neither of these special cases is of direct relevance to the escape of wave energy. Moreover, strictly-parallel waves propagating in the direction of decreasing $\omega_p$, the waves remain in the L~mode, and cross the A-mode dispersion curve, $z=z_{\rm A}$, rather than deviating onto this curve, which would be required to allow the wave to escape to infinity.

Specifically, it is not possible for a parallel-propagating wave to move (from where it is generated at $z'=\infty$, $z=-1/\beta_{\rm s}$) along the portion of the dispersion curve that follows $\omega=\omega_{\rm L}(z)$ to $z=-1$ or to $z=-z_{\rm A}$. For a wave to escape, there must be a cross-field gradient in the source region that causes the ray to refract across field lines.

\subsection{Qualitative discussion of escape of O~mode radiation}
 
The escape of radiation (in the O~mode) from the pulsar magnetosphere requires that it follow a ray path to a region where $\omega_{\rm L}(z)$ is arbitrarily small. Along the ray path, the frequency $\omega'$ is assumed to remain constant, with the dispersion relation in ${\cal K}'$ given by (\ref{drO2}), and with the dispersion relation in ${\cal K}$ given by (\ref{drO1}). By inspection of (\ref{drO2}) with (\ref{drO1}), it is apparent that these two conditions can be satisfied only if the denominator in (\ref{drO1}) becomes sufficiently small as the ratio $\omega_{\rm L}^2(z)/\omega^2$ becomes small. This requires that $z^2-z_{\rm A}^2-\tan^2\theta$ becomes small, which may be rewritten as $-z^2(n^2-1-1/z^2\beta_{\rm A}^2)\approx0$, where $\beta_{\rm A}$ is the Alfv\'en velocity divided by $c$. In this case, using (\ref{drO1}), one finds that the refractive index of the O~mode may be approximated by
\be
n^2\approx1-\frac{\omega_{\rm L}^2(z)\sin^2\theta}{\omega^2}+\frac{\cos^2\theta}{\beta_{\rm A}^2}.
\label{drO3}
\ee 
Thus the escape requirement that the wave propagate along a ray path to a region with $\omega^2\gg\omega_{\rm L}^2(z)$, corresponds to the refractive index approaching unity, with $n^2\to1$ in ${\cal K}$ also implying $n'^2\to1$ in ${\cal K}'$.\footnote{The formal requirement that the refractive index for a wave in any medium must approach unity from below for $\omega\to\infty$ is not necessarily satisfied here for $\theta=0$; there is no inconsistency because our formulae for the wave dispersion are derived in the limit $\omega\ll\Omega_e$, precluding $\omega\to\infty$.} The final term in (\ref{drO3}) gives the dispersion relation for Alfv\'en waves in the limit $\sin^2\theta\to0$, but is unimportant for obliquely propagating waves, and is neglected in the following discussion.

We separate the propagation leading to escape into two parts: near the source, and over large distances. 
 
A cross-field gradient is essential in the source region to allow the waves to refract away from parallel propagation. A cross-field gradient leads to refraction away from parallel propagation, and this allows the decrease in $\omega_{\rm p}^2$ to be balanced by an increase in frequency due to nonzero $\theta$. This is possible because the dispersion relation (\ref{drO1}) implies that $\omega^2/\omega_{\rm p}^2$ increases as $\tan^2\theta$ increases for superluminal waves, $z^2>1$. Consider an SAO generated in a locally overdense region: a cross-field gradient allows the wave to become oblique and to propagate towards a lower-density region. We assume that this occurs and the approximate dispersion relation (\ref{drO3}) applies after this initial stage in the propagation.
 
A qualitative understanding of how escape occurs on a large scale may be based on Snell's law: waves refract towards the direction of increasing refractive index. Provided that the approximate dispersion relation (\ref{drO3}) applies, the propagation is analogous to more familiar cases, such as the escape of radiation generated near the plasma frequency in the solar corona. 
 
\subsection{Cross-field gradient}
 
A  plasma generated by pair cascades in a pulsar magnetosphere is likely to be highly inhomogeneous,  structured both along and across magnetic field lines.  Suppose that pairs are created in ``clouds'' each of which is in a narrow range of cross-field displacement $x$. In such a model, there are large gradients $\partial\omega_{\rm p}/\partial x$ on the scale of an individual cloud. One expects little mixing of plasma across field lines to reduce these gradients. 
 
The curvature drift can lead to some mixing, as may be understood as follows. The curvature of the field lines causes charges to drift across field lines at $v_c$, with
\be
\frac{v_c}{c}=\beta^2\gamma\frac{c}{R_c\Omega_e}\approx0.7\times10^{-6}\gamma\left(\frac{r}{r_{\rm L}}\right)^{5/2},
\label{vc}
\ee
where $R_c$ is the radius of curvature. Because $v_c$ is proportional to $\gamma$, the drift speed is different for different $\gamma$s. This dependence leads to a $\gamma$-dependent gradient across field lines, with the lowest energy particles nearest to the initial field line and the highest energy particles furthest from it. If these relative drifts were to cause charges to separate by a distance larger than the separation between clouds, these effects would tend to smooth out cross-field inhomogeneities. This drift is small for $r\ll r_{\rm L}$, and we assume that it does not smooth out the large cross-field gradients in the pulsar plasma.\footnote{We note that this mixing effect increases with increasing $r$, which is equivalent to decreasing $\omega$ here. In principle, such mixing decreases the efficiency of escape with increasing $r$, or increasing $\omega$, implying that $\mu(\omega)$ is an increasing function of $\omega$, the significance of which is discussed above.}

\subsection{Emission in dense fibers}
 
Consider a model in which the density of pairs  is largest in elongated regions along field lines, which we call fibers, and lower in the region between fibers. (This is similar to a model proposed for type~I solar radio bursts \citep{BS77}.) Let the density within a fiber vary across field lines over a characteristic length $L_{\rm x}$: $\partial\omega_{\rm p}/\partial x=-\omega_{\rm p}/2L_{\rm x}$. The gradient along the field lines, $\partial\omega_{\rm p}/\partial s=-\omega_{\rm p}/2L_s$, is assumed to be much smaller. Specifically, for a dipolar model, one has $L_s\approx2r/3$. We assume $L_{\rm x}\ll r$.
 
With these assumptions, equation (\ref{HE5}) for $\dd\theta'/\dd s'$ may be approximated by 
\be
\frac{\dd\theta'}{\dd s'}\approx\frac{z'\cos\theta'}{2L_{\rm x}},
\label{HE6}
\ee
with $z'>1$ for the waves of interest. Equation (\ref{HE6}) implies that $\theta'$ increases such that the waves refract away from the overdense region, in which they are assumed to be generated, towards the underdense region between fibers. Once a ray is in an underdense region, the region acts as a duct with rays refracting away from overdense edges towards the density minimum. (A similar ducting models models was proposed \citet{D79} for escape of fundamental plasma emission from the solar corona.) With low-density regions between fibers acting as ducts, an O~mode wave is guided outward until the density in the surrounding region is too low for the ducting to continue to be effective. Thereafter the ray path may be approximated by a straight line.
 
\subsection{Escape only from the polar-cap region}
 
It is conventional to distinguish between the polar cap regions, in which the plasma needs to be continually replaced through pair creation, and the closed-field region, in which there is no such requirement. It follows that the foregoing ducting model is plausible only on open field lines. In the absence of strongly field-aligned density structures, the SAOs cannot escape. We assume that this is the case in the closed-field regions. Although the field ${\tilde E}_\parallel$, as given by (\ref{Epp4}) for example, applies in both the open- and closed-field regions, we assume that the fraction of the wave energy in SAOs that can escape is non-zero only in the open-field regions.

\section{Discussion}
\label{sect:discussion}
 
The pulsar radio emission mechanism suggested here overcomes the major difficulties (discussed in Paper~1) with existing mechanisms (CCE, RPE and ADE) that have resonant wave growth as an essential ingredient. Resonance is possible only for subluminal waves, and the dispersive properties of pulsar plasma imply such severe constraints on the wave growth that we regard these mechanisms as untenable, as least as generic mechanisms for all pulsar emission. The alternative we explore is based on the oscillations that arise naturally \citep{Letal05,BT07A,BT07,T10b,TA13}, as the plasma attempts to screen the time-varying parallel inductive electric field $E_\parallel$ associated with an obliquely rotating magnetic dipole. Large-amplitude oscillation (LAOs) lead to acceleration of high-energy particles that trigger pair cascades, populating the magnetosphere with plasma. Our suggestion is that smaller-amplitude versions of these oscillations (SAOs), regarded as waves in the pulsar plasma, occur more widely in the magnetosphere, and that some of the energy in the SAOs can escape directly as the observed radio emission. 

{\br
We comment on the difference between the emission mechanism proposed here and the mechanism recently proposed by \cite{2020arXiv200102236P}.  Common features are that both mechanisms appeal to the intrinsic time-dependence of the problem, both involve cross-field inhomogeneities in the plasma, both assume incomplete screening of the electric field  and both result in predominantly O~mode emission. \cite{2020arXiv200102236P} presented numerical results for the parallel and perpendicular components for the resulting fluctuating electric field, and argued that the transverse component is predominantly in the O~mode, with a broad frequency spectrum. In contrast, we assume that the incomplete screening of the parallel component of the inductive electric field acts as a source for L~mode waves in the pulsar plasma. By analogy with a nonrelativistic counterpart of the problem, in which the fluctuations may be regarded as finite-amplitude Langmuir waves, at the plasma frequency, in a pulsar plasma, the fluctuations in a pulsar plasma may be regarded as finite-amplitude L~mode waves at a specific frequency, denoted by $\omega_0$, and we argue that L~mode waves lie on a single the LO-mode dispersion relation, allowing them to escape directly provided an appropriate escape path is available.

}
 
An implication of our model is that the frequency of the observed emission is equal to the frequency, $\omega_0$, of the SAOs in the pulsar frame ${\cal K}'$ in the source region. The identification of $\omega_0$, estimated in (\ref{omega02}), as the observed frequency of pulsar radio emission is an important implication of the model. The argument for this value of $\omega_0$ is as follows.
The excitation of the SAOs implies that they are nearly temporal oscillations. Although purely temporal oscillations in the rest frame ${\cal K}$ of the plasma have a frequency $\omega_{\rm x}\approx\omega_p/\langle\gamma\rangle^{1/2}$, the driver is in the pulsar frame, ${\cal K}'$, and purely temporal oscillations in this frame have a frequency near $\omega_1$ in ${\cal K}$ and hence near $\omega_1/\gamma_{\rm s}$ in ${\cal K}'$, implying $\omega_0\approx(2\langle\gamma\rangle)^{1/2}\omega_{\rm p}/\gamma_{\rm s}$ in ${\cal K}'$. {\br(This estimate applies for $\rho=1/\langle\gamma\rangle\ll1$; for the largest value, $\rho=1$, considered plausible, one finds $\omega_0=1.6430\,\omega_{\rm p}/\gamma_{\rm s}$ in ${\cal K}'$.)}

The prediction that the frequency of pulsar radio emission has the specific form (\ref{omega02}) has several implications. 
\begin{enumerate}
\item Equations (\ref{omega02a}) and (\ref{omega02}) imply a radius-to-frequency mapping of the form $\omega/2\pi\propto r^{-3/2}$. 
\item In Figure~\ref{fig:omega_0} we plot the height $r/R_*$ versus the streaming Lorentz factor $\gamma_{\rm s}$ for the emission, in the range  $ 0.1\,{\rm GHz} \lesssim \omega_0/2\pi \lesssim 5\,{\rm GHz} $, for various values of $\langle\gamma\rangle$ and $P$. The plots show that the height of emission increases with increasing $P$, consistent with the dependence of height with $P$ inferred from observation \citep{Jetal08}.

\item The plots in Figure~\ref{fig:omega_0} allow one to identify favorable cases for the model to account for the observed emission; for example, modest values of $\langle\gamma\rangle$ and of the ratio $\gamma_{\rm s}/\langle\gamma\rangle$ are generally favored.

\item {\br The fluctuating electric field, ${\tilde E}_\parallel$, is proportional to the parallel component, $E_\parallel$, of the inductive electric field, which is proportional to $\sin\alpha$. This mechanism applies only to an oblique rotator, $\sin\alpha\ne0$, and one would expect the radio power to decrease with decreasing $\sin\alpha$. However, this argument assumes that the fraction ${\tilde E}_\parallel/E_\parallel$ is not a sensitive function of $\sin\alpha$, and this may not be the case.

}

\item  In principle, the dependence on pulsar parameters in (\ref{omega02}) coupled with statistical data for a sufficiently large sample of pulsars may allow constraints on the parameters $\kappa$, $\gamma_{\rm s}$, $\langle\gamma\rangle$ to be inferred from the radio data.

\end{enumerate}

In Section~\ref{sect:Jcor} we develop a semi-quantitative model for the amplitude of the SAOs, based on what we refer to as current starvation. The idea is that the screening of the inductive $E_\parallel$ cannot be perfect because of a phase delay between the unscreened $E_{\rm ind\parallel}$ and the screening field $E_{\rm pot\parallel}$ needed to maintain the time-dependent part of the corotation charge density at its instantaneous value. We calculate the residual ${\tilde E}_\parallel$ in (\ref{Epp1})--(\ref{Epp4}), and suggest that it be identified with the maximum amplitude of the SAOs, such that the maximum power per unit area is $\varepsilon_0|{\tilde E}_\parallel|^2c/2$. In an idealized model in which the dominant emission is from an annular region just inside the last closed field line, this model implies a frequency spectrum for the emitted radiation $I(\omega)\propto\mu(\omega)\omega^{-7/3}$, where $\mu(\omega)$ is the fraction of the power in the SAOs that can escape.
 
We discuss the escape in Section~\ref{sect:escape}. The SAOs are assumed to be generated as parallel-propagating L~waves. A novel feature of dispersion in a pulsar plasma that L-mode waves are on the same dispersion curve as O~mode waves and hence, in principle, L-mode waves may evolve in to O-mode waves, which can escape, without involving any nonlinear or mode-coupling process. However, refraction away from parallel propagation is essential to allow escape. We suggest a model in which the plasma is inhomogeneous across field lines, such that the density contours are nearly field-aligned. Near the source SAOs generated in locally overdense regions, which we refer to as fibers, get refracted towards lower-density regions, and thereafter are ducted outwards with the fibers acting as walls to the ducts. We further argue that these fibers are to be expected in plasma produced through pair cascades, and that they are confined to the open-field regions. We argue that the absence of such fibers in the closed-field regions explains the absence of radio emission from these regions. The dependence of ${\tilde E}_\parallel$ on $\theta_{\rm b}$ then favors emission just inside the last closed field line. We also argue that mixing across field lines, due to the curvature drift, may reduce the cross-field gradient with increasing $r$, implying that $\mu(\omega)$ is an increasing function of $\omega$, and hence that the predicted spectrum of the emission is flatter than the idealized case, $I(\omega)\propto\omega^{-7/2}$, when $\mu(\omega)$ is neglected.
 
The model predicts that the escaping radiation is in the O~mode. It is well established that the observed polarization, at least in some pulsars (with elliptical polarization or orthogonal modes) is strongly affected by propagation effects. The observed polarization is assumed to be characteristic of wave properties in a so-called polarization limiting region, beyond which the plasma is ineffective in further modifying the polarization.  In such a model, observed elliptical polarization is attributed to cyclotron effects in the polarization limiting region, such that the polarization on escape can be substantially different from that at the point of emission \citep[e.g.,][]{BP12}. 
 
 We suggest that the proposed mechanism is a favorable candidate as the generic radio emission mechanism for all pulsars. However, there are various aspects of the emission mechanism that require further investigation and development. We argue that the proposed emission mechanism is most favorable for modest values of the ratio $\gamma_{\rm s}/\langle\gamma\rangle$, and the conditions on the pair cascades for this to be the case need to be explored. A specific weakness in our model is the lack of a quantitative treatment of the probability, $\mu(\omega)$, of escape of the energy in SAOs from any point in the magnetosphere.

\section{Conclusions}
\label{sect:conclusions}
 
We suggest that pulsar radio emission is generated as a consequence of oscillations set up as the plasma attempts to screen a residual parallel component of the (inductive) electric field due to the obliquely rotating magnetic dipole. The oscillations are superluminal L~mode waves that become O~mode waves as they propagate outwards. A notable prediction of this model is the frequency of the waves, which is given by equation (\ref{omega02}), is equal to the frequency of the observed emission. This explicit form for the frequency gives a specific formula for frequency-to-radius mapping, allowing this and other features to be compared with observation.

\section*{Acknowledgements}
We thank Andrey Timokhin for pointing out the relevance of the recent paper \citep{2020arXiv200102236P}. We thank Mike Wheatland, Andrew Melatos and an anonymous referee for helpful comments on the manuscript. The research reported in this paper was supported by the Australian Research Council through grant DP160102932.

\section*{Data Availability}
No new data were generated or analysed in support of this research.
 
\bibliographystyle{mnras}
 
\bibliography{Pulsar_radio_Refs}

\appendix

\section{Fields due to a rotating magnetic dipole}
\label{sect:A}
 
{\br An explicit form for fields around an obliquely rotating magnetized star in vacuo were derived by \citep[e.g.][]{D55}, and for a point dipole at the center of the star, these may be derived from the vector potential, ${\bi A}(t,{\bi x})=\curl({\bi m}(t-r/c)/r)$, where $t-r/c$ is the retarded time, with $d{\bi m}(t)/dt=\bomega_*\times{\bi m}(t)$. In the case of a corotating magnetosphere, the electric field is $\bm{E}_{\rm cor}=-(\bm{\omega}_*\times\bm{x})\times\bm{B}$. Other relevant fields follow from Maxwell's equations. }

In terms of spherical polar coordinates $r,\theta,\phi$ and unit vectors $\hat{\bm{r}},\hat{\bm{\theta}},\hat{\bm{\phi}}$ and the rotational phase $\psi=\omega_*t$, these fields are

\begin{equation}\label{dmf6a}
\begin{split}
    \bm{B}_{\rm dip}
        & = \!\begin{multlined}[t]
                {\mu_0m\over4\pi r^3}\, \{2[\cos\alpha\cos\theta + \sin\alpha\sin\theta\cos(\phi - \psi)]\hat{\bm{r}}\\
                +[\cos\alpha\sin\theta - \sin\alpha\cos\theta\cos(\phi - \psi)]\hat{\bm{\theta}}\\ +\sin\alpha\sin(\phi - \psi)\hat{\bm{\phi}}\},
        \end{multlined}\\
    \bm{E}_{\rm ind}
        & = -{\mu_0\omega_*m\sin\alpha\over4\pi r^2}\, \left[\cos(\phi-\psi)\hat{\bm{\theta}}-\cos\theta\sin(\phi-\psi)\hat{\bm{\phi}}\right],\\
    \bm{E}_{\rm cor} 
        & = \!\begin{multlined}[t]
            -{\mu_0\omega_*m\over4\pi r^2}\sin\theta\{[\cos\alpha\sin\theta-\sin\alpha\cos\theta\cos(\phi-\psi)]\,\hat{\bm{r}}\\
            -2[\cos\alpha\cos\theta+\sin\alpha\sin\theta\cos(\phi-\psi)]\,\hat{\bm{\theta}}\},
        \end{multlined}\\
    \Phi_{\rm cor} 
        & = {\mu_0\omega_*m\over4\pi r}\sin\theta[\cos\alpha\sin\theta-\sin\alpha\cos\theta\cos(\phi-\psi)],\\
    \rho_{\rm cor} 
        & = \!\begin{multlined}[t]
            -{2\omega_*m\over4\pi r^3c^2} [\cos\alpha(3\cos^2\theta-1)\\
            + 3\sin\alpha\sin\theta\cos\theta\cos(\phi-\psi)],
        \end{multlined}\\
    \bm{J}_{\rm cor} 
        & = \!\begin{multlined}[t]
            -{\omega_*^2m\sin\alpha\over4\pi r^2c^2}\,[\sin\theta\cos\theta\sin(\phi-\psi)\,\hat{\bm{r}}\\
            -(\cos^2\theta-\sin^2\theta)\sin(\phi-\psi)\,\hat{\bm{\theta}}\\
            -\cos\theta\cos(\phi-\psi)\,\hat{\bm{\phi}}].
    \end{multlined}
\end{split}
\end{equation}
The parallel component of $\bm{E}_{\rm ind}$ and $\bm{J}_{\rm cor}$ are
\begin{equation}\label{Jcorp1}
\begin{split}
    E_{\rm ind\parallel}
        & = -{\mu_0\omega_*m\sin\alpha\over4\pi r^2\Theta(\theta_{\rm b})}\,[\cos\alpha\sin\theta\cos(\phi - \psi) - \sin\alpha\cos\theta],\\
    J_{\rm cor\parallel}
        & = {\omega_*^2m\over4\pi r^2c^2\Theta(\theta_{\rm b})}\, \sin\alpha\cos\alpha\sin\theta\sin(\phi - \psi),
\end{split}
\end{equation}
with $\Theta(\theta_{\rm b})=\{1+3[\cos\alpha\cos\theta+\sin\alpha\sin\theta\cos(\phi-\psi)]^2\}^{1/2}$.
 
\section{Dispersion for superluminal waves $z^2>1$}
\label{sect:superluminal}
 
We summarize some analytic approximations that apply for superluminal waves, $z^2>1$, based on the RPDF $z^2W(z)$ \citep[e.g.,][]{MG99}, evaluated in the rest frame of a 1D J\"uttner distribution with $\langle\gamma\rangle\gg1$. We write
\be
    z^2W(z)
        = \left\langle\frac{z^2(z^2+\beta^2)}{\gamma^3(z^2-\beta^2)^2}\right\rangle
        \approx z^2(z^2+1)f(z^2,\aV{\gamma})
\label{RPDF0}
\ee
where we assume $ z^2 + 1 \gg 1/\gamma^2 $ and define
\be
f(z^2,\langle\gamma\rangle)=\left\langle\frac{1}{\gamma^3(z^2-1+1/\gamma^2)^2}\right\rangle.
\label{RPDF1}
\ee
The wave properties of interest are the dispersion relation $\omega=\omega_{\rm p}[z^2W(z)]^{1/2}$, the ratio of electric to total energy,
\be
R_{\rm L}(z)=
-\frac{W(z)}{zdW(z)/dz}\approx-\frac{(z^2+1)f(z^2,\langle\gamma\rangle)}{z[2zf(z^2,\langle\gamma\rangle)+(z^2+1)f'(z^2,\langle\gamma\rangle)]},
\label{app4}
\ee
with $f'(z^2,\langle\gamma\rangle)=\partial f(z^2,\langle\gamma\rangle)/\partial z$, and the group speed,
\be
\beta_{\rm g}(z)=\frac{d[z^2W(z)]/dz}{z\,dW(z)/dz}=z[1-2R_{\rm L}(z)].
\label{betag}
\ee
 
For $z^2\gg1$ the only relevant average is $\langle\gamma^{-3}\rangle\approx1/\langle\gamma\rangle$, and one finds
\be
f(z^2,\langle\gamma\rangle)\approx\frac{1}{(z^2-1)^2\langle\gamma\rangle},
\quad
f'(z^2,\langle\gamma\rangle)\approx\frac{-4z}{(z^2-1)^3\langle\gamma\rangle}.
\label{RPDF2}
\ee
These give
\be
R_{\rm L}(z)\approx\frac{z^4-1}{2z^2(z^2+3)}\approx\frac{1}{2}\left(1-\frac{3}{z^2}\right),
\label{z2gg1a}
\ee
where the second approximation applies for $z^2\gg3$, and
\be
    \beta_{\rm g}(z)
        \approx\frac{3z^2+1}{z(z^2+3)}
        \approx\frac{3}{z}\left(1-\frac{8}{3z^2}\right),
\label{z2gg1b}
\ee
where the second approximation applies for $z^2\gg3$. These approximations generalize the wave properties for $z^2=\infty$, corresponding to strictly temporal oscillations in ${\cal K}$, to finite $z^2\gg1$.
 
For strictly temporal oscillations in ${\cal K}'$ one has $z'^2=\infty$, $z^2=1/\beta_{\rm s}^2\approx1+1/\gamma_{\rm s}^2$. In this case we need $f(z^2,\langle\gamma\rangle)$ and $f'(z^2,\langle\gamma\rangle)$ at $z^2=1/\beta_{\rm s}^2\approx1+1/\gamma_{\rm s}^2$:
\bea
    f(1/\beta_{\rm s}^2,\langle\gamma\rangle)
        &\approx& \left\langle\frac{\gamma}{(1+\gamma^2/\gamma_{\rm s}^2)^2}\right\rangle,\nn\\
    f'(1/\beta_{\rm s}^2,\langle\gamma\rangle)
        &\approx& -4\left\langle\frac{\gamma^3}{(1+\gamma^2/\gamma_{\rm s}^2)^3}\right\rangle.
\label{RPDFa}
\eea
For $z^2=1$ one has $f(1,\langle\gamma\rangle)=\langle\gamma\rangle$,
$f'(1,\langle\gamma\rangle)=-4\langle\gamma^3\rangle=-24\langle\gamma\rangle^3$, where we use $ \av{\gamma^2} \approx n!\av{\gamma}^n $ for $ \av{\gamma} \gg 1 $, and hence
\be
R_{\rm L}(z)\approx\frac{1}{24\langle\gamma\rangle^2},
\qquad
\beta_{\rm g}\approx1-\frac{1}{12\langle\gamma\rangle^2},
\label{app5}
\ee
to lowest order in an expansion in $1/\langle\gamma\rangle^2$. The limit $z^2\to1$ corresponds to the minimum value of $R_{\rm L}$ and the maximum value of $\beta_{\rm g}$ for $z^2\ge1$.  (We note that although this limit corresponds to retaining only the leading term in an expansion in $\gamma^2/\gamma_{\rm s}^2$ of the denominators in (\ref{RPDF1}), this expansion appears not to converge due to $n$th term in the expansion being proportional to $(-)^n(n+1)\langle\gamma^{2n+1}\rangle/\gamma_{\rm s}^{2n}$ and $(-)^n(n+1)(n+2)\langle\gamma^{2n+3}\rangle/2\gamma_{\rm s}^{2n}$, respectively, with $\langle\gamma^{2n+1}\rangle=(2n+1)!\langle\gamma\rangle^{2n+1}$,  $\langle\gamma^{2n+3}\rangle=(2n+3)!\langle\gamma\rangle^{2n+3}$, cf.\ Appendix~D of RMM1.
 
 The group speed in ${\cal K}'$ is
\be
\beta'_{\rm g} 
    = \frac{\beta_{\rm s} + \beta_{\rm g}}{1 + \beta_{\rm s}\beta_{\rm g}} 
    \approx 1 - \frac{1 + 3\av{\gamma}^2/\gamma_{\rm s}^2}{48\gamma_{\rm s}^2\av{\gamma}^2},
\label{betagp}
\ee
where we use~\eqref{app4} so that $ \gamma_{\rm g}^2 \approx 6\av{\gamma}^2 $.

{\br
\subsection{Numerical values for $\rho=1$}
\label{rho=1}
Analytic expressions for arbitrary $\rho$ are given by \cite{MG99}  in their Table~1:
\be
\langle\gamma\rangle=\frac{K_2(\rho)+K_0(\rho)}{2K_1(\rho)},
\quad
\langle\gamma^{-1}\rangle=\frac{K_0(\rho)}{K_1(\rho)},
\quad
\langle\gamma^{-3}\rangle=\frac{Ki_2(\rho)}{K_1(\rho)},
\label{MG991}
\ee
where $K_n(\rho)$ is a Macdonald function and $Ki_n(\rho)$ is the $n$th integral of $K_0(\rho)$. Numerical values for $\rho=1$ give
\be
\langle\gamma\rangle=1.69948,
\quad
\langle\gamma^{-1}\rangle=0.69948,
\quad
\langle\gamma^{-3}\rangle=0.45459.
\label{MG991}
\ee
These values give $\omega_1^2=2.69948\omega_p^2$ and $\omega_x^2=0.45459\omega_p^2$ for $\rho=1$.
}
\end{document}